\documentclass[pre,epsfig,address,twocolumn]{revtex4}
\usepackage[english]{babel}
\usepackage{amsmath}
\usepackage{amssymb}
\usepackage{color}
\usepackage{graphicx}
\usepackage{bm}
\usepackage{cleveref}
\usepackage{times,amsmath,amssymb}
\usepackage{amsfonts}
\usepackage{amssymb}
\usepackage{color}
\usepackage{textgreek}

\usepackage{algorithmicx}

\newcommand{\be}{\begin{equation}}
\newcommand{\ee}{\end{equation}}
\newcommand{\bel}[1]{\begin{equation}\label{#1} } 
\newcommand{\eeascsacas}{\end{equation}}

\begin{document}
\title{Mirror symmetry breakdown in the Kardar–Parisi–Zhang universality class} 

\author{Johannes Schmidt$^{1,2}$, Andreas Schadschneider$^2$}
\affiliation{
$^{1}$Bonacci GmbH, 50937 Cologne, Germany.\\
$^{2}$Institut f\"{u}r Theoretische Physik, Universit\"{a}t zu K\"{o}ln, 
50937 Cologne, Germany.\\
}

\begin{abstract}
The current/height fluctuation statistics of Kardar-Parisi-Zhang (KPZ) universality in 1+1 dimensions are sensitive to the initial state.
We find that the averages over the initial states exhibit universal and scale-invariant patterns when conditioning on fluctuations.
To establish universality of our findings we demonstrate scale invariance at different times and heights using large-scale Monte-Carlo simulations of the totally asymmetric simple exclusion process (TASEP) which belongs to the KPZ universality class.
Here we focus on current/height fluctuations in the steady state regime described by the Baik-Rains distribution.
The conditioned probability distribution of an initial state order parameter shows a transition from uni- to bimodal. Bimodality occurs for negative current/height fluctuations that are dominated by super-diffusive shock dynamics. It is caused by two possible point-symmetric shock profiles and the KPZ mirror symmetry breakdown.
Similar surprising relations between initial states and fluctuations might exist in other universality classes as well.
\end{abstract}

\date{\today }
\maketitle


\section{Introduction}
Anomalous and exotic types of dynamical laws have been at the forefront in recent years.
Understanding the relationship between initial states and fluctuations of conserved currents is essential for diagnosing dynamic phenomena in and out of equilibrium.
Besides the wide interest in the Kardar-Parisi-Zhang (KPZ) universality class, fluctuations are non-Gaussian, super-diffusive and sensitive to initial states.
In this paper, we show for the KPZ universality class a pervasive mirror symmetry breakdown.
By establishing scaling laws for probability distributions and conditioned observables we demonstrate universality and provide a new strategy to analyse dynamical structures.
One of the seminal models in this class is the Totally Asymmetric
Simple Exclusion Process (TASEP) and its relatives \cite{ASEP1,ASEP2,ASEP3}. The TASEP has been extensively
studied both in physics and mathematics and by now many of its properties are rather
well understood. It has been realized that its behavior is paradigmatic for a large
class of driven diffusive systems that typically belong to the Kardar-Parisi-Zhang (KPZ)
universality class \cite{KPZ,Halp15}. The KPZ class has been found to be rather robust, e.g. generic
generalizations of the TASEP to multilane situations or longer-ranged hopping as in
the Nagel-Schreckenberg model of traffic flow belong to the same class \cite{GSSS_PRE_2019}.
In the following we will take a closer look at the importance of initial states for current/height fluctuations in the steady-state regime.
This way new insights are obtained, e.g. about the possible relation between fluctuations and the occurrence of symmetry breaking.
The results are important not only for the particular class of systems studied here as they provide us with new tools for understanding transport phenomena with non-Gaussian fluctuations in and out of equilibrium \cite{Krajnik_22_PRL, Krajnik_23_PRX, Krajnik_23_PRL}.

\begin{figure}
\centering{}\includegraphics[width=8.3cm]{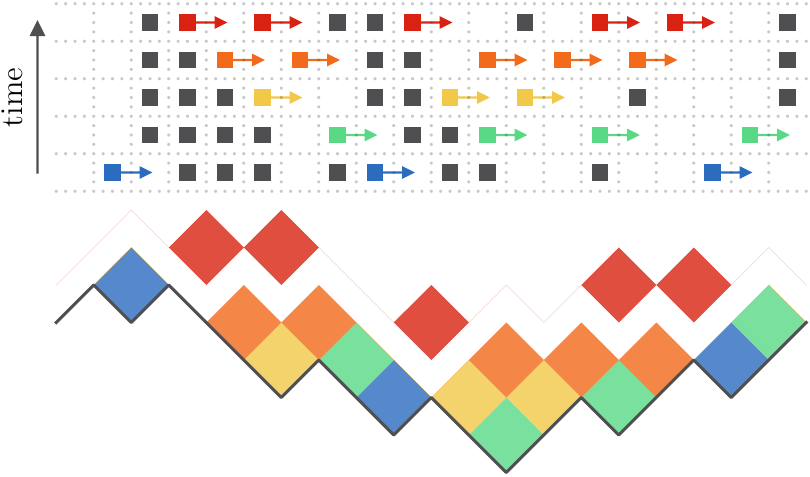}
\caption{\label{fig:tasep_kpz_map} Mapping the TASEP dynamics to surface growth. Shown is a TASEP configuration 
evolving in time. The colored particles will hop at the next time step. Mapping a particle to a down-slope ($\blacksquare\rightarrow\diagdown$) and a hole to an up-slope ($\square\rightarrow\diagup$) one obtains a height 
profile for each TASEP configuration. If a particle hops to the right a diamond is added to the surface between 
the particle's initial and final position.}
\end{figure}


\begin{figure}[t]
\centering{}\includegraphics[width=8.5cm]{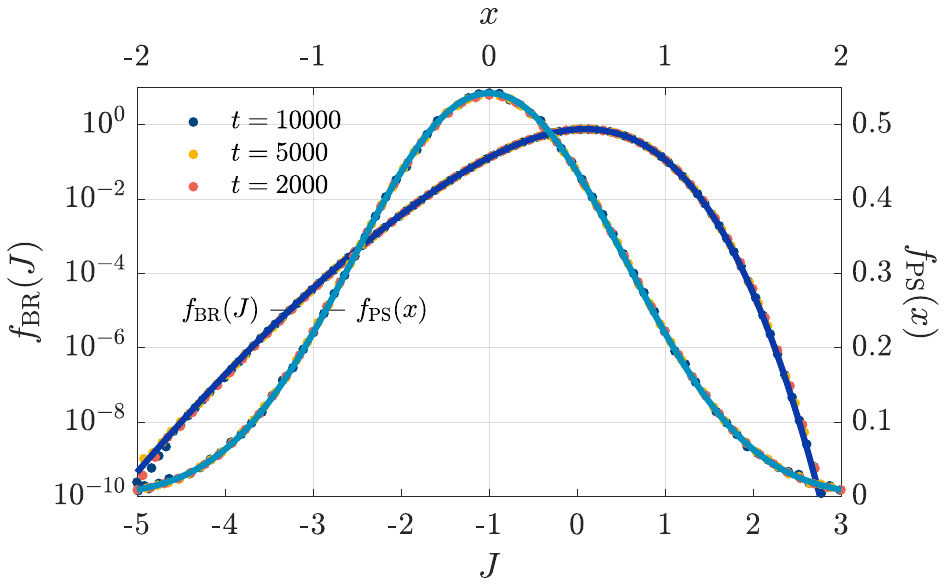}
\caption{\label{fig:BR_and_PS}
Baik-Rains distribution in log-scale (left bottom axes) and Prähofer-Spohn scaling function (right top axes).
The data shown were recorded in a TASEP with hopping probability $p=1/2$. The re-scaling has been made according to Eqs.~(\ref{eq:Baik_Rains_Current_Dis}), (\ref{eq:struc_fct_asymp_scaling}) and shows a nice agreement to their asymptotic forms.
The scaling functions $f_{\mathrm{BR}}(J)$ and $f_{\mathrm{PS}}(J)$ are tabulated with high precision in \cite{Prae04_DATA} where $f_{\mathrm{BR}}(J)=2F^{\prime}_{0}(-2J)$ with $F_{0}(\cdot)$ defined in \cite{Prae04_DATA}.
}
\end{figure}

\section{Theory}
In recent years it has been shown that the dynamical properties of driven diffusive lattice gases,
not only those in the KPZ universality class \cite{Fibo1,Fibo2}, are rather well described by their hydrodynamic density field $\mathtt{u}(\text{x},t)=\varrho(\text{x},t)-\rho$ which is the deviation
of the local density field $\varrho(\text{x},t)$ from its stationary background $\rho$. Based on the conservation law $\partial_{t}\varrho(\text{x},t)+\partial_{\text{x}}\mathtt{j}(\text{x},t)=0$ and steady-state current-density relation $j(\rho)$ the non-linear fluctuating hydrodynamic (NLFH) equation describes the evolution of $\mathtt{u}(\text{x},t)$ \cite{GSSS_PRE_2019} as 
\begin{equation}
\partial_t \mathtt{u} = -\partial_\text{x} \left( \text{v} \mathtt{u}
                     +\frac{\lambda}{2}\mathtt{u}^2
                     - \nu\partial_\text{x} \mathtt{u} + \sqrt{B} \eta
                     \right)
\label{eq:NLFH_equation}
\end{equation}
with the collective velocity $\text{v}=j^{\prime}(\rho)$, non-linearity $\lambda=j^{\prime\prime}(\rho)$, diffusion constant $\nu$ and space-time white noise $\eta$ of strength $B$.
After a Galilean transformation $\tilde{a}(\tilde{\text{x}},t)=a(\text{x},t)$ with $\tilde{\text{x}}=\text{x} - \text{v}t$ removing the drift term $-\text{v}\partial_{\text{x}}\mathtt{u}$
and introducing a new variable $\mathtt{\tilde{h}}(\tilde{\text{x}},t)$ by
\begin{equation}
\partial_{\tilde{\text{x}}}\tilde{\mathtt{h}}(\tilde{\text{x}},t)=-\mathtt{u}(\tilde{\text{x}}+\text{v}t,t)
\end{equation}
the NLFH equation
turns into the KPZ equation \cite{KPZ}
\begin{equation}
\partial_{t}\tilde{\mathtt{h}}=\nu\partial_{\tilde{\text{x}}}^{2}\tilde{\mathtt{h}}+\frac{\lambda}{2}\left(\partial_{\tilde{\text{x}}}\tilde{\mathtt{h}}\right)^{2}+\sqrt{B}\tilde{\eta} \,
\label{eq:KPZoriginal}
\end{equation}
where $\mathtt{\tilde{h}}(\tilde{\text{x}},t)$ is a surface height profile growing at average speed $j(\rho)$.
Note the mirror symmetry of the KPZ equation under the transformation $\tilde{x}\rightarrow -\tilde{x}$.
For the lattice model the substitution $\partial_{\tilde{\text{x}}}\tilde{\mathtt{h}}=-\tilde{\mathtt{u}}$
is motivated by the exact mapping of the TASEP to a discrete surface
growth process \cite{BARABASI_GROWTH,HHZ,KRUG_GROWTH,KRUG_Book_GROWTH,TASEP_GROWTH}
that is known as the single-step model (Fig.~\ref{fig:tasep_kpz_map}).
The universal large-scale properties for {\em typical} fluctuations of the KPZ equation are by now
well-understood, see \cite{Halp15} for a review.
The dynamical exponent that relates the scaling of space and time
variables as $\text{x}\sim t^{1/z}$ takes the value $z=3/2$, as opposed
to $z=2$ for normal
diffusion or $z=1$ of the deterministic Eulerian scaling.

Our goal is to study the interplay between typical fluctuations of current/height and the underlying 
initial states within the steady state regime.
Time-integrated currents or height fluctuations are given as
\begin{align}
\mathtt{J}_t &=\int_{0}^{t}\left[\mathtt{j}(\text{v} t,s)
-j(\rho)\right]\mathrm{d}s
- \int_{0}^{\text{v}t}\mathtt{u}
\left(\text{x},0\right)\mathrm{d}\text{x}
\label{timeint_current}
\\
& = \tilde{\mathtt{h}}(0,t)-\tilde{\mathtt{h}}(0,0) - j(\rho)t
\label{height-fluctuatuon}
\end{align}
and its fluctuations are of general interest \cite{QuastelSpohn2015, Spohn_LesHouches_2015, Takeuchi2018}.
A prominent exact result is the Baik-Rains (BR) distribution \cite{BR} for the steady state regime
\begin{equation}
\mathcal{P}_{\mathrm{BR}}(\mathtt{J}_t)\simeq(\Gamma t)^{-\frac{1}{3}}f_{\mathrm{BR}}\left(\mathtt{J}_t\cdot(\Gamma t)^{-\frac{1}{3}}\right)\label{eq:Baik_Rains_Current_Dis}
\end{equation}
with scaling parameter $\Gamma =4|\lambda|\nu^2/B^2$ (Fig.~\ref{fig:BR_and_PS}).
The initial state region around the measure point that contributes to typical current/height fluctuations
can be specified by the dynamical structure function $S(\text{x},t)$ \cite{Prae04}. At large times it has the form
\begin{align}
S(\text{x},t)&\equiv \frac{B}{\nu}\left<\mathtt{u}(\text{x},t)\mathtt{u}(0,0)\right>\nonumber\\
&\simeq(Et)^{-\frac{2}{3}}f_{\mathrm{PS}}\left((Et)^{-\frac{2}{3}}(\text{x}-\text{v}t)\right)
\label{eq:struc_fct_asymp_scaling}
\end{align}
with scaling parameter $E = |\lambda|\sqrt{2\nu/B}$ (Fig.~\ref{fig:BR_and_PS}).
From this scaling behavior, we infer a significance length $\xi_{\alpha,t} = c_\alpha(Et)^{2/3}$.
Here $c_\alpha$ is a significance factor that ensures a $1-\alpha$ confidence level that current/height fluctuations are dominated by the initial state within a radius $\xi_{\alpha,t}$ around the measure point. The confidence level is determined through $1-\alpha = \intop_{-c_\alpha}^{c_\alpha}f_{\text{PS}}(x)\mathrm{d}x$.
The scaling functions $f_{\mathrm{BR}}$ and $f_{\mathrm{PS}}$ (Fig. 2) have analytical expressions that can not be expressed in terms of most usual special functions. They are tabulated with high precision in \cite{Prae04_DATA}\footnote{$f_{\mathrm{BR}}(J)=2F^\prime_0(-2J)$ with $F_0(\cdot)$ defined in \cite{Prae04_DATA}}.

\begin{figure}[t]
~~\phantom{a}
~~\phantom{a}
\includegraphics[width=8.25cm]{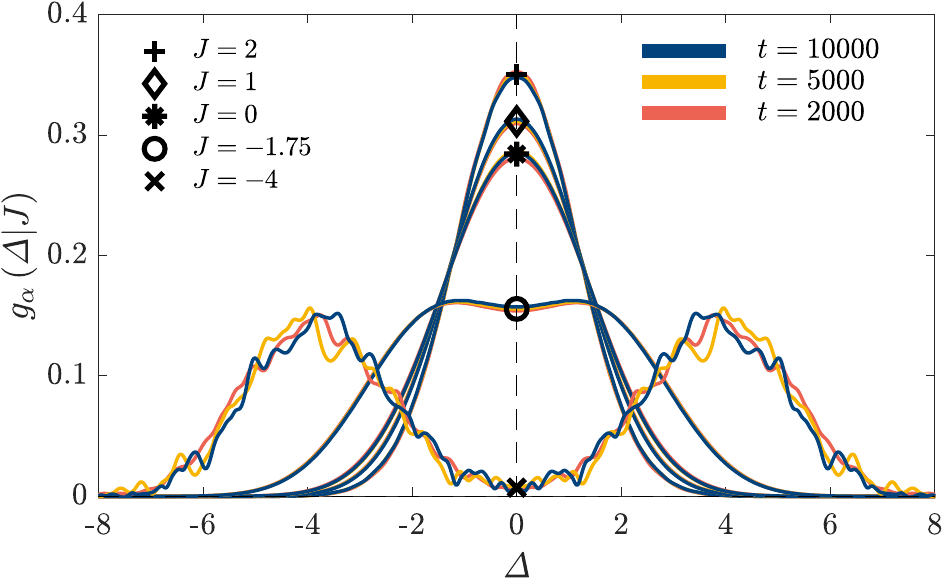}
\caption{\label{fig:Delta_Distribution}
Distributions for the re-scaled order parameter $\varDelta$ at different times conditioned on different re-scaled currents $J$.
The data shown were recorded in a TASEP with hopping probability $p=1/2$. The re-scaling has been made according to Eq.~(\ref{Eq:Delta_Distribution}) 
and shows a nice collapse.
}
\end{figure}


\section{Simulations}

To numerically investigate KPZ universality class in the steady-state regime the TASEP model with parallel update rule and periodic boundary conditions offers analytical and computational advantages \cite{GSSS_PRE_2019,MeerSchm2017}.
The TASEP model is defined on a one-dimensional uniform lattice of size $L$. Each lattice site can be occupied by a maximum of one particle. For the parallel update rule the time is uniformly discretized. In one time step, the particles jump to the right with probability $p$ given that the neighboring site is free (Fig.~\ref{fig:tasep_kpz_map}).
Initial states are drawn according to the steady state measure \cite{SSNI_1995}
\begin{equation}
    \mathcal{P}\left(\{n_{\text{x},0}\}\right)=\prod_{\text{x}=1}^{L}P_{n_{\text{x},0},n_{\text{x}+1,0}}
\end{equation}
with 
\begin{align}
P_{1,0}&=\frac{1}{2p}\sqrt{1-4p\rho(1-\rho)}\\
P_{1,0}=P_{0,1};~P_{1,1}&=\rho-P_{1,0};~P_{0,0}=1-\rho-P_{1,0}
\label{eq:stat_measure}
\end{align}
where $n_{\text{x},t}\in\{0,1\}$ is the occupation number of site $\text{x}$ at time $t$ and $\rho = \left<n_{\text{x},t}\right>$ the average density. 
The resulting steady-state current-density and fluctuation-dissipation relation are
\begin{align}
j(\rho) & =\frac{1}{2}\left[1-\sqrt{1-4p\rho(1-\rho)}\right]\\
\nu/B & =\rho(1-\rho)\sqrt{1-4p\rho(1-\rho)}. \label{eq:stat_curr_paralel}
\end{align}
Note that Eqs.~(\ref{eq:stat_measure})-(\ref{eq:stat_curr_paralel}) are only valid for TASEP with parallel update rule. Knowing the model's steady state measure allows to construct initial states directly and avoid numerically expensive relaxations.
Due to the fastest convergence into the asymptotic scaling regime and by the particle hole symmetry improved statistics all simulations are performed with $\rho=1/2$ \cite{GSSS_PRE_2019}. To validate that the model enters the asymptotic regime, described by the KPZ universality, we test for the scaling relations Eqs.~(\ref{eq:Baik_Rains_Current_Dis}) and (\ref{eq:struc_fct_asymp_scaling}), see Fig.~\ref{fig:BR_and_PS}.
To optimize the memory usage the TASEP's state is encoded bitwise into a $\mathtt{int32}$ datatype. Further, limiting the hopping probabilities to $p\in\{\frac{1}{2},\frac{1}{4}\}$ allows to use the $\mathtt{int32}$ random numbers bitwise and propagate the model with bitwise operations.
This efficiency gain allows to increase the system size and observation times. Larger systems allow to suppress finite-size effects of order $\mathcal{O}(L^{-1})$ and increase statistics by using the translational invariance.
All states are propagated independently and random numbers are generated using a MT19937 pseudo random number generator.
To get the time integrated current for a discrete model one simply replaces the integrals in Eq.~(\ref{timeint_current}) by sums and uses the discrete instantaneous current $\mathtt{j}_{\text{x},t}=n_{\text{x},t}(1-n_{\text{x},t})n_{\text{x+1},t+1}$ and fluctuation $u_{\text{x},t}=n_{\text{x},t}-\rho$ fields.
Monte-Carlo simulations are performed with systems of length $L=2\cdot 10^9$. For the system with $p=1/2$ ($p=1/4$) we used $705$ ($1305$) independent realizations.
In order to treat the regions around the current measure point as independent, we demand that the measure points are at least at least $3\xi_{\alpha,t}$ apart. This results in at least $40$ initial state samples for $J=-4$  and up to $2.8\cdot10^8$ for $J=0$. The number of expected initial state samples for a given current $J$ and time $t$ can be roughly estimated using the Baik-Rains distribution Eq.~(\ref{eq:Baik_Rains_Current_Dis}) and the system size.


\section{Results}
To focus on individual initial states, we use the order parameter defined in \cite{MeersonKamenev2016},
which measures the initial mass or height difference around the measure point,
but modified with a scaled significance cutoff $\xi_{\alpha,t}$, i.e. 
\begin{align}
\Delta &=-\int_{-\xi_{\alpha,t}}^{\xi_{\alpha,t}}\mathtt{u}\left(\text{x},0\right)\text{dx}
\\
&=\tilde{\mathtt{h}}\left(\xi_{\alpha,t},0\right) - \tilde{\mathtt{h}}\left(-\xi_{\alpha,t},0\right).
\end{align}
Space correlation of $\mathtt{u}$ decay exponentially, so the central limit theorem reveals $\Delta$ to be Gaussian distributed.
By mirror symmetry one has $\left\langle \Delta\right\rangle =0$ and using the stationary fluctuation dissipation relation $\nu/B=\int_{-\infty}^{\infty}\left<\mathtt{u}(0,t)\mathtt{u}(\text{x},t)\right>\text{d}x$
one finds for the variance $\left\langle \Delta^{2}\right\rangle \simeq c_\alpha (\Gamma t)^{2/3}$. Therefore, the distribution of the order parameter full fills a simple scaling law 
\begin{align}
\mathcal{P}_{\alpha}\left(\Delta\right)&\simeq\left(\Gamma t\right)^{-1/3}g_{\alpha}\left(\left(\Gamma t\right)^{-1/3}\Delta\right)\\
g_{\alpha}\left(\varDelta\right)&=\frac{1}{\sqrt{2\pi c_{\alpha}}}\exp\left(-\frac{\varDelta^{2}}{2c_{\alpha}}\right).
\end{align}
where $g_{\alpha}$ solely depends on the significance factor $c_{\alpha}$. As $\xi_{\alpha,t}$ grows logarithmically slow in $\alpha$ we expect our study to be qualitatively independent of $\alpha$ and limit our observations to $c_\alpha=2$, leading to a significance factor $\alpha<0.4\%$.
To analyze the role of initial states given a time integrated current, we investigate the probability distribution of the order parameter given a measured current, i.e. $\mathcal{P}_\alpha\left( \Delta|{\mathtt{J}_t} \right)$.
Due to missing theoretical prediction we measure $\mathcal{P}_\alpha\left( \Delta|{\mathtt{J}_t} \right)$ via TASEP data and establish its scaling property as
\begin{align}
\mathcal{P}_\alpha\left( \Delta|{\mathtt{J}_t} \right) \simeq
\left(\Gamma t\right)^{-\frac 1 3}
g_{\alpha}\left(\left.\left(\Gamma t\right)^{-\frac 1 3}\Delta\right|\left(\Gamma t\right)^{-\frac 1 3}\mathtt{J}_t\right)
\label{Eq:Delta_Distribution}
\end{align}
where $g_{\alpha}\left(\varDelta\mid J\right)$ is an universal conditioned scaling function (Fig.~\ref{fig:Delta_Distribution}). The conditioned distribution $g_{\alpha}\left(\varDelta\mid J\right)$ is related to $g_\alpha(\varDelta)$ through the law of total probability, i.e.
\begin{equation}
\int_{-\infty}^{\infty}g_{\alpha}\left(\varDelta\mid J\right)f_{\text{BR}}\left(J\right)\text{d}J=g_{\alpha}\left(\varDelta\right).
\end{equation}
Remarkably, $g_{\alpha}\left(\varDelta\mid J\right)$
reveal for $J<0$ a bi-modal and $J>0$ an uni-modal structure.
The bi-modality indicates a breakdown of the KPZ mirror symmetry within the slow decaying tail ($J<0$) of the BR distribution.
For a decreasing current we observe a continuous transition from a uni- to bi-modal distribution resulting in two well separated peaks at $\pm J$, where the resolution of the transition to bi-modality is sensitive $\alpha$.
A similar effect, a bi-modality with two sharp peaks at $\pm J$, has been predicted for large deviations in the early time regime for $J>J_c$ \cite{MeersonKamenev2016, MeersonKamenev2018} and is supported by a recent numerical study \cite{HartmannMeerson21}.
Different to typical fluctuations, large deviations scale as $\mathtt{J}_t\sim t$, and the significance length grows as $\xi\sim t$.
Here we stress that the effect of bi-modality or symmetry breaking appears in the asymptotic regime for negative fluctuations of typical scale $\mathtt{J}_t\sim t^{1/3}$.  Therefore we argue that symmetry breaking plays an important role in general and is likely to be observed in experimental settings.

\begin{figure}[t]
\centering{}\includegraphics[width=8.25cm]{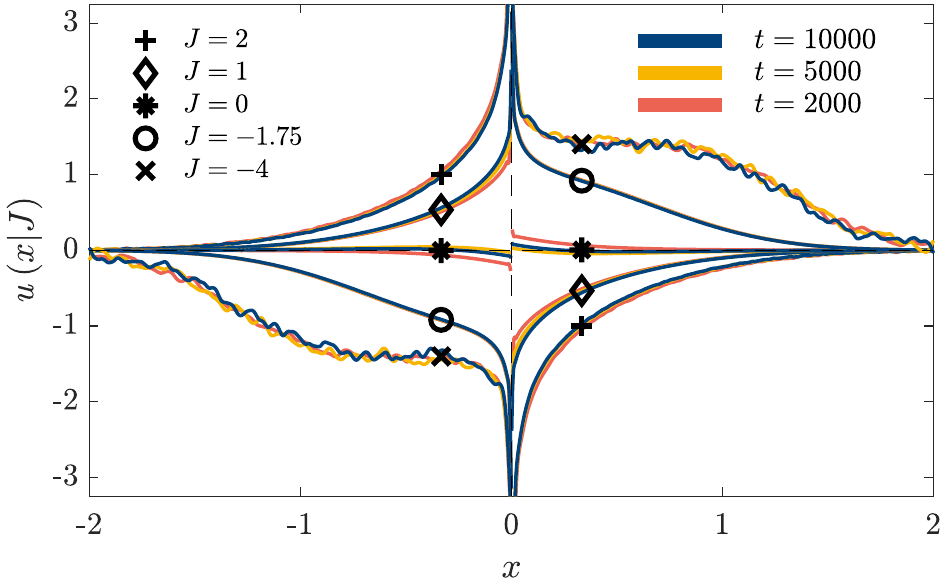}
\caption{\label{fig:collapsed_density_profiles}
Space-rescaled initial density profiles conditioned on the rescaled current. The figure shows how deviations from the expected current impact the initial density profiles.
The re-scaling is according to Eq.~(\ref{Eq:u_scaling_function}) and shows a nice collapse. The data shown were recorded in a TASEP with hopping probability $p=1/2$.
}
\end{figure}

The order parameter distribution shows that fluctuations are linked to specific initial states. 
Therefore, we quantify the expected density profiles around the measure point when conditioning on a realized current/height.
From TASEP data (Fig.~\ref{fig:collapsed_density_profiles}) we find the universal scaling property
\begin{align}
\left\langle \left.\mathtt{u}\left(\text{x},0\right)\right|_{\mathtt{J}_t}\right\rangle \simeq 
\left(\frac{\Gamma}{tE^{2}}\right)^{\frac 1 3}\mathit{u}\left(\left.\left(Et\right)^{-\frac 2 3}\text{x}\right|\left(\Gamma t\right)^{-\frac 1 3}\mathtt{J}_t\right)
\label{Eq:u_scaling_function}
\end{align}
with
\begin{equation}
\int_{-\infty}^{\infty}\left|\mathit{u}\left(x|J\right)\right|\mathrm{d}x  \simeq J\label{Eq:mu-mass_identity}
\end{equation}
and limited by the support of the $f_{\text{PS}}$ scaling function.
This suggests that initial states realizing a current fluctuation of strength $J$ show on average an absolute height change or absolute mass difference of $J$.
To extract the two possible expected density profiles in the mirror symmetry broken regime, we limit our investigation to $J=-4$ where the distribution $g_\alpha(\Delta|J)$ has two well separated peaks. Using Eq.~(\ref{Eq:u_scaling_function}), but additionally conditioning on $\Delta$, the symmetry broken profiles are obtained via
\begin{align}
\label{Eq:u_pm_calculation}
\mathit{u}_{\pm}\left(\left.x \right|J \right) = \mathit{u}\left(x \left|J, \pm\Delta > 0\right.\right).
\end{align}
and show a nice collapse for two different TASEP systems at different times underlying the universality (Fig.~\ref{fig:collapsed_density_profiles}).
The KPZ mirror symmetry now is reflected in $u_{+}(x|J)=-u_{-}(-x|J)$ and we find $\intop_{-\infty}^{\infty}\mathit{u}_{\pm}\left(x|J\right)\mathrm{d}x \simeq \pm J$.
The symmetry broken initial state profiles turn out to be super-diffusive shocks.
Different to a typical shock \cite{shock1,shock2} the size scales as $\sim (E t)^{2/3}$ and its velocity as $\sim \left( \frac{\Gamma}{E^2t} \right)^{1/3}$.
This means the size of shock and its within time $t$ moved distance are similar to the spread of fluctuations.
We stress that these conditioned density profiles appear after averaging over many initial states.
The distribution of the lack of mass Eq.~(\ref{Eq:Delta_Distribution}) indicates how strong a single realization may deviate from the expected profile.
This is in contrast to the large deviation regime \cite{HartmannMeerson21}, where a single realization is close to the expected profile.

\begin{figure}[t]
\centering{}\includegraphics[width=8.25cm]{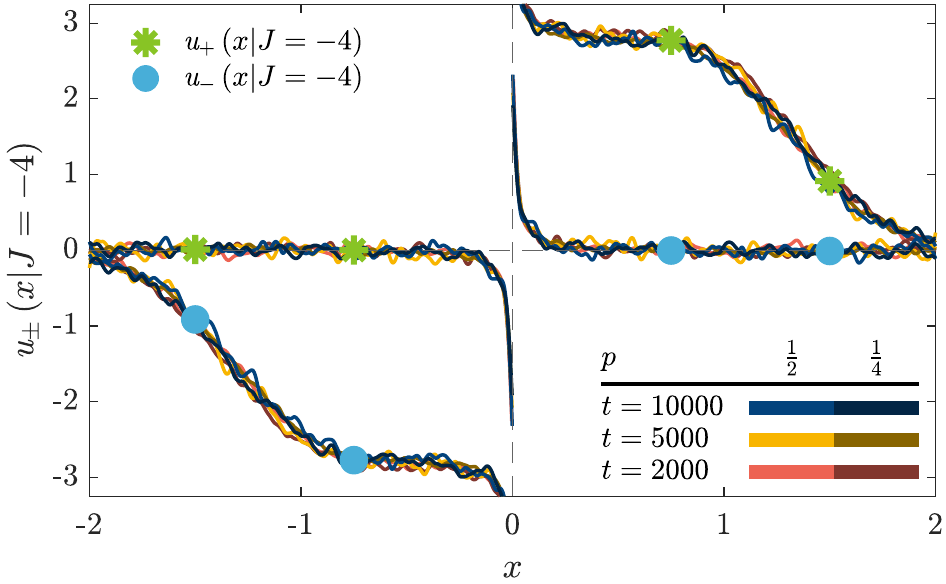}
\caption{\label{fig:broken_sym_profile}
Re-scaled mirror symmetry broken initial density profiles.
The re-scaling is according to Eqs.~(\ref{Eq:u_scaling_function}), (\ref{Eq:u_pm_calculation}) and shows a nice collapse.
The data shown were recorded in a TASEP with hopping probabilities $p=1/2$ and $p=1/4$.
}
\end{figure}


\section{Conclusions}
The results presented here shed new light on the relation between initial states and the statistics of fluctuations.
They generalize previous findings and put them into a new perspective, e.g. the connection with a breaking of the KPZ mirror symmetry for negative current fluctuations.
Furthermore, universal scaling properties for conditioned observables have been established which shows the generality of the results and could be useful for further studies.

Although we have considered only the KPZ universality class in the steady-state regime and its representation by the TASEP we point out that the results are far more general. It is remarkable, that even though given a Gaussian distributed order parameter, conditioning on a non-Gaussian current fluctuation reveals a bi-modal distribution that indicates a symmetry breaking.
Similar connections can be made even for systems with non-Gaussian fluctuations belonging to other universality classes in and out of equilibrium \cite{Krajnik_22_PRL,Krajnik_23_PRX, Krajnik_23_PRL}, explicitly pointed out in \cite{Krajnik_23_PRX}.
The observed results are related to the fact that small currents can be realized at two different densities. In the ASEP this is related to the particle-hole symmetry. It would be interesting to investigate what happens in models like the variant of the Katz-Lebowitz-Spohn model studied in \cite{HagerKPS2001}. Here the same current can be realized at up to four different densities which might lead to multi-modal distributions.

\section*{Acknowledgements}
We are very grateful to J. Krug, B. Meerson, V. Pasquier and V. Popkov for encouragement and useful comments. We thank anonymous referees for useful suggestions.
JS gratefully acknowledges the hospitality of the Racah Institute of Physics, Hebrew University of Jerusalem, where the inspiration for this paper came from lively discussions with B. Meerson.
Simulations have been performed using the HPC cluster Cheops of the University of Cologne.


\begin{thebibliography}{99}

\bibitem{ASEP1}
J.T. MacDonald, J.H. Gibbs, and A.C. Pipkin, Biopolymers 6, 1 (1968).

\bibitem{ASEP2}
B. Derrida, Phys. Rep. 301, 65 (1998).

\bibitem{ASEP3}
G.M. Sch\"utz  Phase Transitions and Critical Phenomena,
Vol. 19 (Academic Press, London, 2001).

\bibitem{KPZ}
M. Kardar, G. Parisi, and Y.-C. Zhang, Phys. Rev. Lett. {56}, 889 (1986).

\bibitem{Halp15}
T. Halpin-Healy and K.A. Takeuchi,
J. Stat. Phys. 160, 794 (2015).

\bibitem{GSSS_PRE_2019}
J. de Gier, A. Schadschneider, J. Schmidt, G.M. Sch\"utz,
Phys. Rev. E 100, 052111 (2019)

\bibitem{Krajnik_22_PRL}
Ž. Krajnik, J. Schmidt, V. Pasquier, E. Ilievski, T. Prosen,
Phys. Rev. Lett. 128, 160601 (2022)

\bibitem{Krajnik_23_PRX}
Ž. Krajnik, J. Schmidt, V. Pasquier, E. Ilievski, T. Prosen,
Phys. Rev. Res. 6, 013260 (2024)

\bibitem{Krajnik_23_PRL}
Ž. Krajnik, J. Schmidt, E. Ilievski, T. Prosen,
Phys. Rev. Lett. 132, 017101 (2024)

\bibitem{Fibo1} 
V. Popkov, A. Schadschneider, J. Schmidt, G.M. Sch\"utz, PNAS 112, 12645 (2015)

\bibitem{Fibo2} 
V. Popkov, A. Schadschneider, J. Schmidt, G.M. Sch\"utz, JSTAT (2016) 093211

\bibitem{HHZ}
T. Halpin-Healy and Y.-C. Zhang, Phys. Rep. {254}, 215 (1995).

\bibitem{KRUG_Book_GROWTH} J. Krug and H. Spohn, \textit{Kinetic
    roughening of growing surfaces} (Cambridge University Press,
  Cambridge, 1991).

\bibitem{BARABASI_GROWTH} A.-L. Barab\'asi and H.E. Stanley,
  \textit{Fractal Concepts in Surface Growth} (Cambridge University
  Press, Cambridge, 1995).

\bibitem{KRUG_GROWTH}
J. Krug and H. Spohn,
Phys. Rev. A {38}, 4271 (1988).

\bibitem{TASEP_GROWTH}
M. Plischke, Z. R\'acz, and D. Liu,
Phys. Rev. B {35}, 3485 (1987).

\bibitem{QuastelSpohn2015}
J. Quastel and H. Spohn, J. Stat. Phys. 160, 965 (2015).

\bibitem{Spohn_LesHouches_2015}
H. Spohn,
\textit{Lecture notes for Les Houches Summer School 2015}, arXiv:1601.00499.

\bibitem{Takeuchi2018}
K. A. Takeuchi, Physica A 504, 77 (2018).

\bibitem{BR} 
J. Baik and E.M. Rains, J. Stat. Phys. {100}, 523 (2000).

\bibitem{Prae04}
M. Pr\"ahofer and H. Spohn,
J. Stat. Phys. {115}, 255 (2004).

\bibitem{Prae04_DATA}
M. Pr\"ahofer and H. Spohn, http://www-m5.ma.tum.de/KPZ

\bibitem{MeersonKamenev2016}
M. Janas, A. Kamenev, B. Meerson, Phys. Rev. E 94, 032133 (2016).

\bibitem{MeersonKamenev2018}
N.R. Smith, A. Kamenev, B. Meerson, Phys. Rev. E 97, 042130 (2018).

\bibitem{HartmannMeerson21}
A.K. Hartmann, B. Meerson, P. Sasorov, Phys. Rev. E 104, 054125 (2021).

\bibitem{shock1}
E.D. Andjel, M.D. Bramson, T.M. Liggett, Probab. Theory Relat. Fields 78, 231 (1988).

\bibitem{shock2}
C. Boldrighini, G. Cosimi, S. Frigio, M. Grasso Nunes, J. Stat. Phys. 55, (1989).

\bibitem{MeerSchm2017}
B. Meerson, J. Schmidt, J. Stat. Mech. 129901 (2017).

\bibitem{SSNI_1995}
M. Schreckenberg, A. Schadschneider, K. Nagel, N. Ito, Phys. Rev. E 51, 2939 (1995).

\bibitem{HagerKPS2001}
J.S. Hager, J. Krug, V. Popkov, G.M. Sch\"utz, Phys. Rev. E 63, 056110 (2001).

\end{thebibliography}
\end{document}